\documentclass[12pt]{article}\usepackage[hyperfootnotes=false]{hyperref}
\usepackage{epsfig}
\usepackage{float}

\usepackage{caption}

\usepackage{amsmath}
\usepackage{amssymb}
\usepackage{graphicx}
\newlength{\abstractwidth}
\renewcommand{\thefootnote}{\fnsymbol{footnote}}
\renewcommand{\thanks}[1]{\footnote{#1}} 
\newcommand{\starttext}{
\setcounter{footnote}{0}
\renewcommand{\thefootnote}{\arabic{footnote}}}
\renewcommand{\theequation}{\thesection.\arabic{equation}}
\newcommand{\be}{\begin{equation}}
\newcommand{\bea}{\begin{eqnarray}}
\newcommand{\eea}{\end{eqnarray}}
\newcommand{\beq}{\begin{equation}}
\newcommand{\ee}{\end{equation}}
\newcommand{\eeq}{\end{equation}}

\def\ba{\begin{eqnarray}}
\def\ea{\end{eqnarray}}

\def\12{{1 \over 2}}

\def\ra{\rangle}

\def\simleq{\; \raise0.3ex\hbox{$<$\kern-0.75em
\raise-1.1ex\hbox{$\sim$}}\; }
\def\simgeq{\; \raise0.3ex\hbox{$>$\kern-0.75em
\raise-1.1ex\hbox{$\sim$}}\; }

\def\O2{\Omega_2}

\def\bi{\begin{itemize}}
\def\ei{\end{itemize}}

\def\sc{\setcounter{equation}{0}}

\def\W{$\Omega$}
\def\W'{$\Omega$}

\def\V{\Omega}
\def\V'{\Omega}

\def\O{{\cal{O}}}

\def\c{{\cal{C}}}

\def\op{opaque}
\def\tr{transparent}
\def\try{transparency}

\def\bn{\bigskip \noindent}

    \def\cg{$\c$-geometry}
     \def\cg2{$\c_2$-geometry}

\makeatletter
\g@addto@macro\normalsize{%
  \setlength\abovedisplayskip{10pt}
  \setlength\belowdisplayskip{20pt}
  \setlength\abovedisplayshortskip{10pt}
  \setlength\belowdisplayshortskip{20pt}
}
\makeatother

\usepackage{color}

\begin{document}
\renewcommand{\theequation}{\thesection.\arabic{equation}}
\begin{titlepage}
\rightline{}
\bigskip
\bigskip\bigskip\bigskip\bigskip
\bigskip

\centerline{\Large \bf {The Typical-State Paradox:}}

\bn

\centerline{\Large \bf { Diagnosing Horizons}}

\bn

\centerline{\Large \bf { with Complexity}}

\bigskip
\begin{center}
\bf   Leonard Susskind  \rm

\bigskip

 Stanford Institute for Theoretical Physics and Department of Physics, \\
Stanford University,
Stanford, CA 94305-4060, USA \\
\bigskip

\end{center}

\begin{abstract}

The concept of transparent and opaque horizons is defined. One example of opaqueness is the presence of a firewall. Two apparently contradictory statements are reconciled: The overwhelming number of black hole states have opaque horizons; and: All black holes formed by natural processes have transparent horizons. A diagnostic is proposed for transparency, namely that the computational complexity of the state be increasing with time. It is shown that opaque horizons are extremely unstable and that the slightest perturbation will make them transparent within a scrambling time.

\medskip
\noindent
\end{abstract}


\end{titlepage}

\starttext \baselineskip=17.63pt \setcounter{footnote}{0}
\tableofcontents

\sc
\section{Introduction}
 I will use the adjective \it transparent \rm to indicate that a horizon can be freely crossed by an in-falling observer. The term refers to the one-way transparency of a future horizon.  The opposite of transparent is opaque. Opacity is a very unstable condition, and as we will see, throwing in a single thermal photon will render the horizon \tr \ in a scrambling time.
 
 \bn
 
  \bn

\begin{figure}[H]
\begin{center}
\includegraphics[scale=.3]{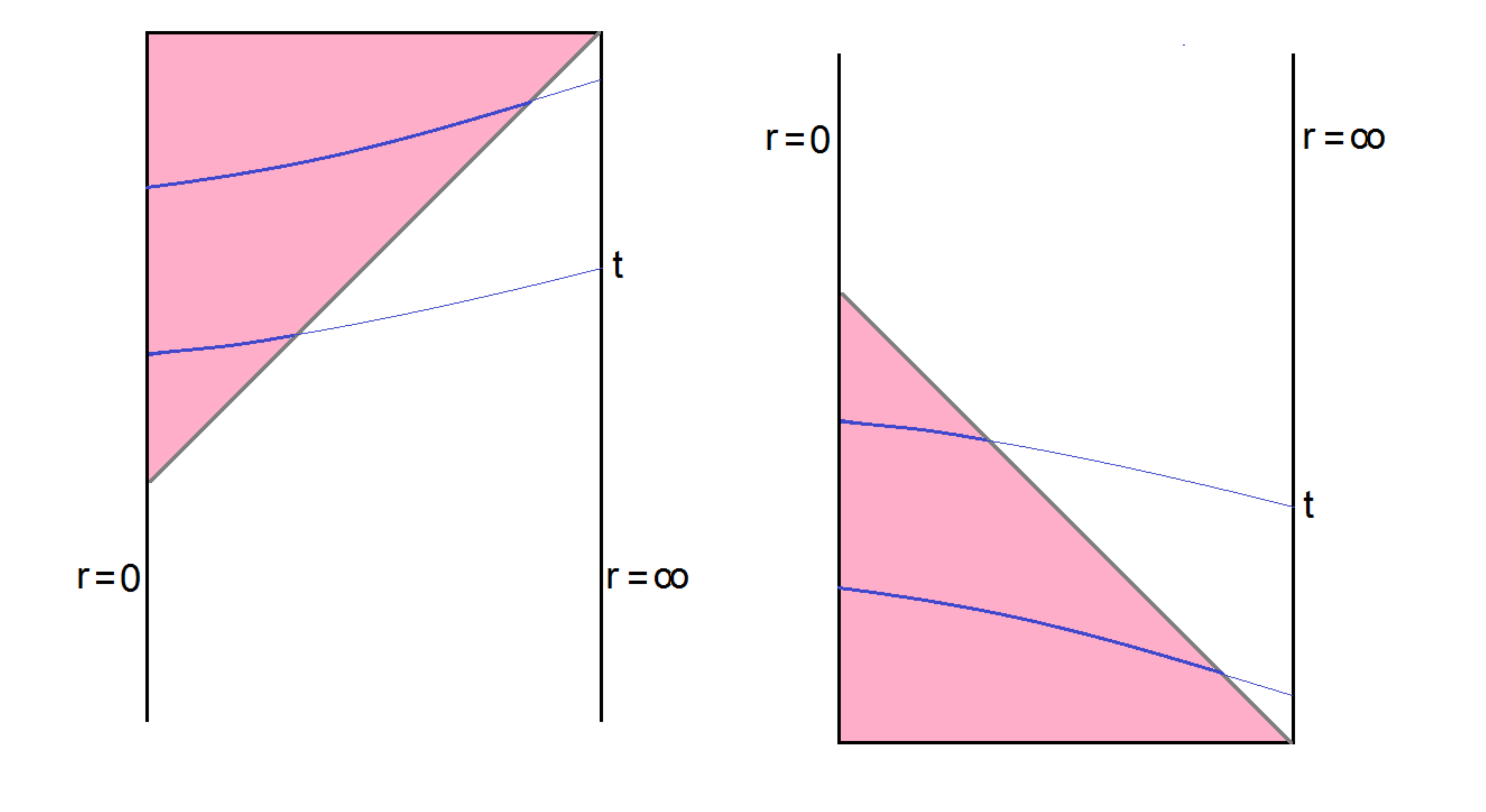}
\caption{The left panel shows a one-sided ADS black hole. The pink region contains the bridge-to-nowhere. The geometry can be foliated by maximal volume slices anchored at the boundary.  The right panel shows a white hole similarly foliated. In the black hole case the volume of the bridge increases with $t$ and in the white hole the volume decreases.}
\label{fig1}
\end{center}
\end{figure}

 
 There are at least two reasons for believing that the horizons of some black holes  may  be opaque. The first is that the black hole in question may not be a black hole at all, but rather a white hole. White holes may not be common in nature but they are just as common as black holes in the Hilbert space describing systems of entropy $S.$  In figure \ref{fig1} a one-sided ADS black hole, and a white hole, are shown. Both spaces are foliated by maximal slices according to the prescription of \cite{Susskind:2014moa}\cite{Susskind:2014jwa}.
  It is evident that the black hole horizon can be entered, but the white hole horizon cannot.
\bn

  The other possible reason for opacity  is that a so-called firewall may seal off the interior geometry  \cite{Almheiri:2012rt}.

 The questions I want to explore are: 

\bi 
\item 
what are the various criteria on the quantum state of the black hole that determine the transparency of the horizon?

\item Is there a single criterion?

\item Do black holes produced by natural processes satisfy the criteria, and if so, for how long a time?

\item How robust are opaque and \tr \ horizons?
\ei

The answers that I propose are that there is a single criterion for transparency, and  that all black holes produced in nature satisfy it. Black holes produced by natural processes in anti de Sitter space are transparent for at least an exponentially long time but  no longer than a doubly exponential time. Black holes in empty flat space will evaporate long before the horizon becomes opaque. Finally transparent horizons are very robust; opaque horizons are very fragile and can be made transparent by a single thermal photon.

\sc
\section{The Typical State Paradox} \label{S typical}

There is  obvious tension between the  three statements:

\bn

1)	The horizons of typical states in the Hilbert space of a black hole are opaque.

2)	Black holes quickly become typical once they reach thermal equilibrium.

3)	Black holes in nature have transparent horizons.

\bn

 Marolf and Polchinski \cite{Marolf:2015dia} argued that typical black hole states (the overwhelming majority) cannot have  transparent horizons. I'm not completely convinced, but in this paper I'll assume that this worse-case scenario  is correct.

The third statement is surely true right after the horizon forms during a collapse process. Take for example a black hole that forms when a light-like shell passes its Schwarzschild radius. The horizon forms before the shell arrives and the space inside the shell is perfectly flat. It's unreasonable in this situation to think that a firewall exists. If we add the fact that the shell can have arbitrarily small energy density as it crosses the Schwarzschild radius, we must conclude that the smooth horizon continues past the shell. Finally, there is no known semi-classical mechanism to create a firewall from a configuration that has a smooth horizon.


The second statement is the problem. It has  two parts to it.  First is that a black hole quickly reaches thermal equilibrium. That much is true;  the time for a solar mass black hole to thermalize  is about a millisecond. After that, all ordinary properties of the black hole as measured from the outside will be very close to their thermal equilibrium values. This led AMPS to claim that by a scrambling time black hole states become typical. The scrambling time   is given by

\be
t_{\ast} =\frac{\beta}{2\pi} \log S
\label{2.1}
\ee
where $\beta$ is the inverse temperature of the black hole and $S$ is its entropy \cite{Hayden:2007cs}\cite{Sekino:2008he}\cite{Maldacena:2015waa}.

The second part of the statement---that thermal equilibrium may be identified with typicality---is the part which is not correct. The resolution of the typical state paradox lies in the fact that
thermal equilibrium is just the tip of an enormous complexity iceberg\footnote{Computational complexity was first introduced into black hole physics by Harlow and Hayden \cite{Harlow:2013tf} in a somewhat different context.} \cite{Susskind:2014rva}\cite{Stanford:2014jda}\cite{Susskind:2014jwa}\cite{Susskind:2014moa}. In classical discrete systems like cellular automata, maximum complexity is  the same as maximum entropy, but in quantum systems there is a huge gulf between maximum entropy and maximum complexity. 
It takes an exponentially long time (exponential in the entropy) to create a typical state of maximum complexity.

To put it another way, there are two kinds of equilibria that we need to distinguish. The first---thermal-equilibrium---occurs when coarse-grained entropy reaches its maximum, and  local observables reach their equilibrium values. This usually happens in a  short time.  The second kind of equilibrium is what I call  complexity-equilibrium.  It occurs when complexity reaches its maximum. 
The time scale to reach  complexity-equilibrium is at least exponential in the  entropy.  Therefore there is a huge stretch of time  during which  a  black hole is in thermal equilibrium, but is still evolving toward complexity equilibrium. States with sub-exponential complexity are extraordinarily  rare, and  not at all  typical.

Let's consider black holes that are created by \it natural processes\rm---processes
 that occur on time scales much less than exponential. These black holes are necessarily  far from maximal complexity, and they remain so  for an exponential time. They belong to the class of non-typical thermal-equilibrium states, which are  far from complexity-equilibrium.

\bn

\sc
\section{Criteria for Transparency}

There are three criteria, any of which should be sufficient to guarantee a transparent horizon. They are not independent; in fact I believe they are all the same. The first   is that the interior geometry of the black hole  be expanding \cite{Susskind:2014rva}. The expansion can be seen from the  Penrose diagram in the left panel of figure \ref{fig1}
where the pink region represents the expanding future geometry of an Einstein-Rosen ``bridge-to-nowhere." In the right panel the pink region indicates a contracting   ERB.  The foliation of the bridge-to-nowhere geometry by maximal space-like surfaces was described in detail in \cite{Susskind:2014jwa}.

\bn
  \bf Expansion Criterion \rm

\it Black hole horizons are \tr \ as long as the interior geometry is expanding. \rm

\bn

The second criterion was suggested by Page \cite{Page:2013mqa}:
Classical black holes with contracting interior volumes will have singularities in their past. This follows from time reversal; black holes with expanding geometries have singularities in the future. This leads to the second criterion:

\bn

\bf Page Criterion   \rm

\it  
Black
 holes formed from non-singular Cauchy data, i.e., no singularities in their past, have transparent horizons. \rm \   

\bn

In fact black holes with singularities in the past are not black holes at all. They are  white holes: statistical freaks\footnote{White holes are analogous to Boltzmann brains, but with complexity replacing entropy. Although I called them statistical freaks, over the infinite stretch of time, assuming ADS boundary conditions, they are as common as black holes. Real black holes evaporate long before they become white holes. In this respect the situation is similar to the cosmological analog of metastable de Sitter space which decays before Boltzmann brains occur.} that are born with high complexity that is fine-tuned to decrease. Nevertheless, I will reluctantly follow common terminology  and call these objects black holes.

The Expansion and Page criteria are expressed in classical geometric terms. They do not directly  refer to  properties of the quantum state. 
The third criterion is  strictly about quantum information. It is best expressed for ADS black/white  holes for which a precise quantum description exists.
It is related to the first criterion by the complexity/volume duality of \cite{Stanford:2014jda}.  According to the C/V duality the volume  (  $V$ ) of the interior of the black hole is proportional to the computational complexity ($\c$) of its quantum state. The precise connection for neutral ADS black holes is, 

\be
\c = \frac{V}{G l_{ads}}
\label{3.1}
\ee

\bn

\bf Increasing Complexity Criterion  \rm

\it  A Black hole horizon will be transparent as long as the the complexity of its quantum state is increasing. \rm

\bn 

On one side of the  C/V duality the complexity is defined for the quantum state of the dual CFT at time $t.$ On the other side the volume is defined for the Einstein-Rosen bridge (or bridge-to-nowhere) on the set of maximal slices anchored at the boundary at time $t.$ Figure \ref{fig1} shows the bridge-to-nowhere foliated by maximal slices. The volume of the slices obviously decreases with time in the white hole and increases in the black hole. It follows that the quantum complexity of the state decreases in the white hole and increases in the black hole.
According to the conjecture of this paper, the decreasing complexity   in the white hole implies that the horizon is opaque. It's obvious from the figure that the past horizon can not be crossed.

\bn

Since in the black hole case complexity can be expected to increase for an exponential time $\sim e^S,$ the horizon will be transparent far beyond the   scrambling or Page times. The other side of the coin is that quantum mechanically, complexity is bounded, implying that the eternal classical expansion of the  geometry must eventually break down. A question of some importance is how long does the complexity of a chaotic system like a black hole continues to grow, and how large does it eventually become? In this paper I will assume that the growth is as rapid as possible for as long as possible. This would mean that complexity grows linearly for a time exponential in the black hole entropy, and then levels off.

In   section \ref{Evidence for ICC}  precise  evidence will be given for the validity of   the Increasing Complexity Criterion for transparency.

\bn
\sc
\section{How Rare is Transparency?}

By the space of states of a black hole  I will mean the space of normalized  pure states that are  macroscopically  indistinguishable, and which from the outside behave like  classical black holes with given macroscopic parameters. The space  has dimensionality $\sim e^S.$ Technically it is the space $CP(e^S).$ 
The resolution of the typical state paradox is that black hole states formed by natural processes are an extremely rare subset of the space of $CP(e^S).$   To be quantitative about how rare natural black holes are, we consider the  history of a chaotic quantum system with many degrees of freedom. The quantum evolution is a trajectory through $CP(e^S),$
restricted to an $e^S$-dimensional torus. To see why this is true let's write the time-dependent state of the system in the energy basis,

\be
|\Psi(t)\ra = \sum_n F_n e^{-i E_n t}|n\ra
\label{4.1}
\ee
where $|n\ra $ is the nth energy eigen-state and $F_n$ are a fixed set of real amplitudes. The entire time dependence is through the phases $e^{-i E_n t}$ which define a torus.
If the system is chaotic the $E_n$ will be incommensurate and the motion on the torus will be ergodic. 
By ergodic I am not referring to  the   behavior of classical systems in phase space.   
Ergodicity  in this context refers to the evolution of the state-vector  in Hilbert space. 

Classically ergodicity implies that
 time-averages equal phase space averages on energy surfaces, \rm 
 
 Ergodicity in Hilbert space implies 
that time averages equal averages over  tori in Hilbert space.\rm

\bn

 Assuming we start with a simple state, the complexity has no choice but to increase. It can be proved that it will increase linearly with time for some period 	\cite{Nielsen}. There are very strong reasons to believe that it will continue to increase until it reaches an exponentially large value. Any stalling of the complexity growth before it becomes exponential in $S$  would imply an unexpected  complexity-class collapse\footnote{ Aaronson and Susskind, to eventually be published. }. 

We'll assume that complexity  increases until it becomes maximal; in other words until  complexity equilibrium has been established.  Once this happens the  complexity  will remain close to  maximal (with fluctuations) for a quantum recurrence time $t_{qr}.$  The quantum recurrence time is doubly  exponential in the entropy,    

\be
\log{   \log{t_{qr}}  } \sim S.
\label{4.2}
\ee

On the scale $t_{qr}$ the state will quasi-periodically return to simple states of low complexity, and then be free to grow again for a singly exponential time.  It follows that the fraction of volume of $CP({e^S})$ that corresponds to small but growing complexity is doubly exponentially small. 
The remaining volume describes states that may have firewalls, or be otherwise opaque.

The reader may be puzzled about why he or she has never heard of complexity equilibrium, ergodicity in Hilbert space, exponential equilibration times, and doubly exponential recurrence times. Even for black holes, we wouldn't be interested in these things if we were only concerned about the outside of the black hole. Complexity is far too subtle  to be relevant for ordinary properties of matter or  the outsides of black holes. It was very surprising to  me that  these unfamiliar, overly subtle  things are directly related to the geometric properties of the space behind  horizons. 

\bn

To summarize, there is no conflict between the claims that typical horizons are opaque, and  that for sub-exponential time natural horizons are transparent (see however section \ref{S-linearity}). Moreover, by a polynomial time real black holes in flat space will have evaporated.

\sc
\section{Evidence for  Criterion}\label{Evidence for ICC}

The example given in section \ref{S typical} is somewhat trivial and does not consider the possible role of firewalls in making the horizon opaque. 
In this section  quantitative evidence is given for the more general validity of the increasing complexity criterion as insurance against firewalls.  

\begin{figure}[H]
\begin{center}
\includegraphics[scale=.5 ]{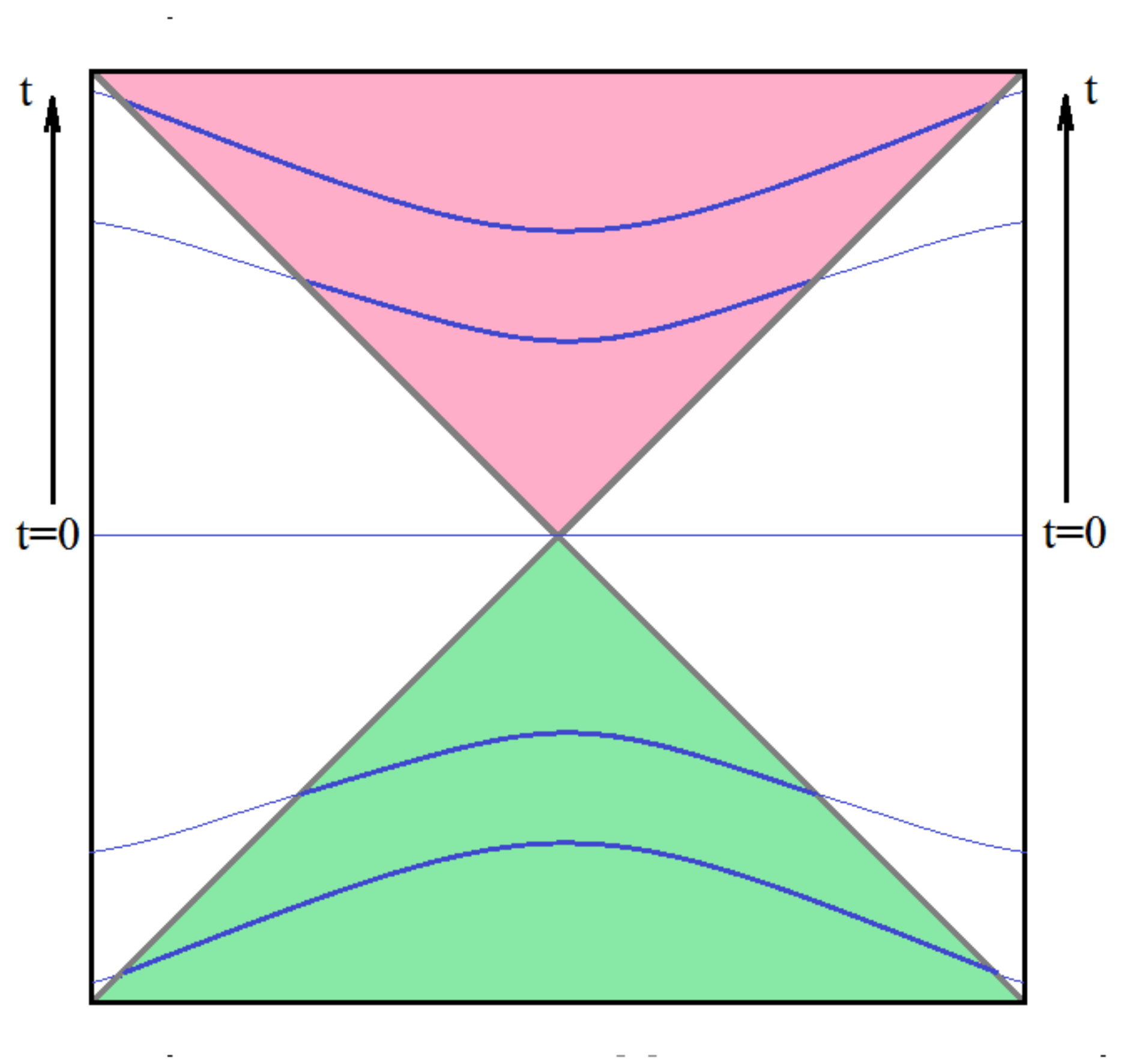}
\caption{Eternal black/white  hole foliated by maximal volume slices anchored on the boundary. The dark blue lines represent the ERB behind the horizon. The volume of the ERB decreases in the white hole region and increases in the black hole region.}
\label{fig2}
\end{center}
\end{figure}

\begin{figure}[H]
\begin{center}
\includegraphics[scale=.5]{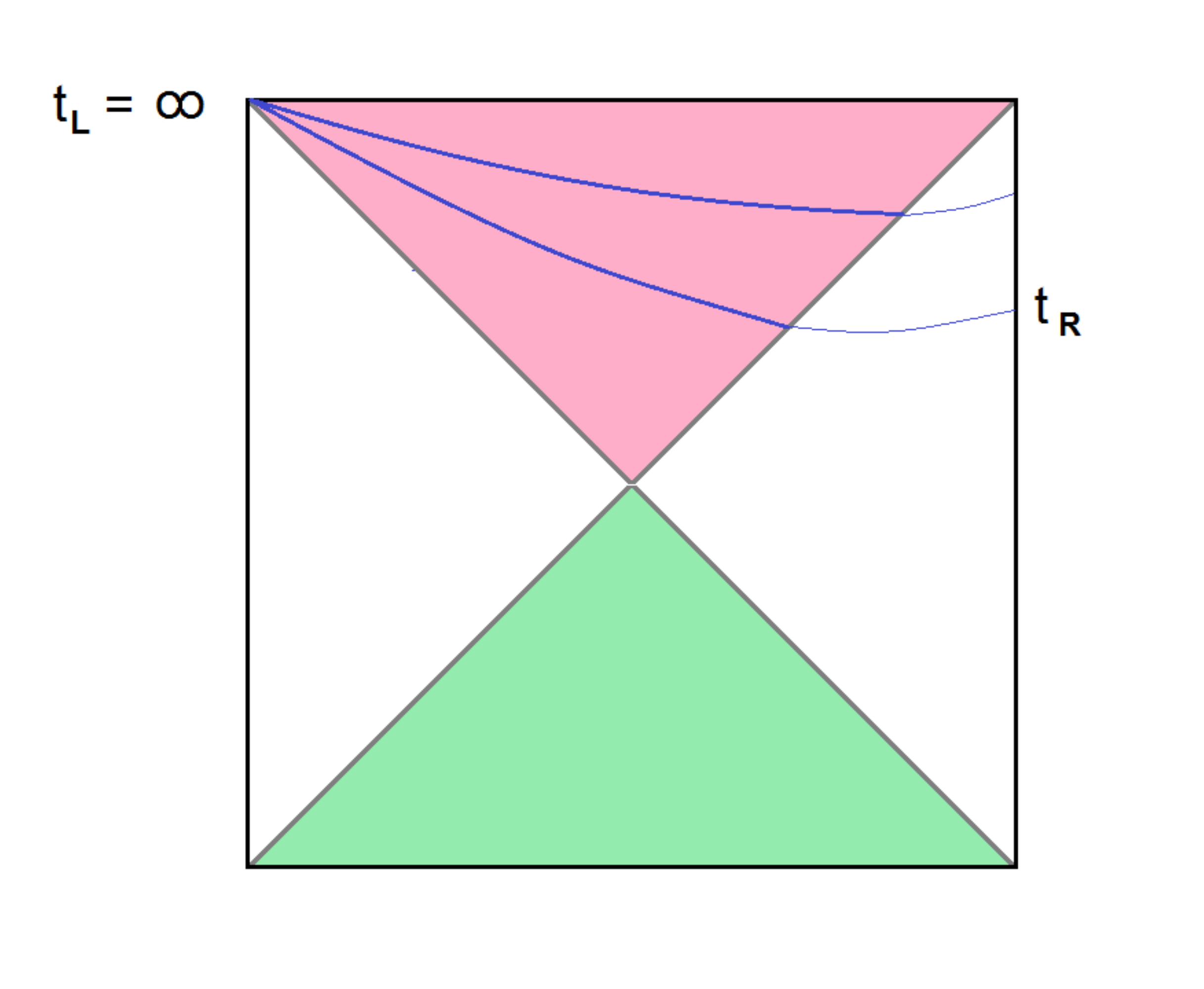}
\caption{To diagnose the right horizon we anchor the maximal slices at $t_L = \infty$ and vary $t_R.$}
\label{fig3}
\end{center}
\end{figure}

The examples in this section come from the theory of two-sided eternal black holes.  For definiteness we will concentrate on a black hole of Schwarzschild radius equal to the ADS length scale $l_{ads}.$
The Penrose diagram, foliated by maximal slices, is shown in figure \ref{fig2}.            
As we will see it is possible to have the horizon on one side be opaque while the other horizon is not. To diagnose one horizon what we must do is to study the complexity/volume as a function of the time on that side. Define $t_L$ and $t_R$ to be the left and right anchoring points. To study the right horizon we will fix $t_L$ and vary $t_R.$
To avoid ambiguities $t_L$ can be chosen in an invariant way by letting it tend to $\infty$ as in 
figure \ref{fig3},  while $t_R$ is  varied. The volume of the ERB and the complexity of the  quantum state  increases with increasing $t_R.$ This means that the right horizon is transparent. Obviously the  same argument can be made on the left.
\bn

\begin{figure}[H]
\begin{center}
\includegraphics[scale=.5]{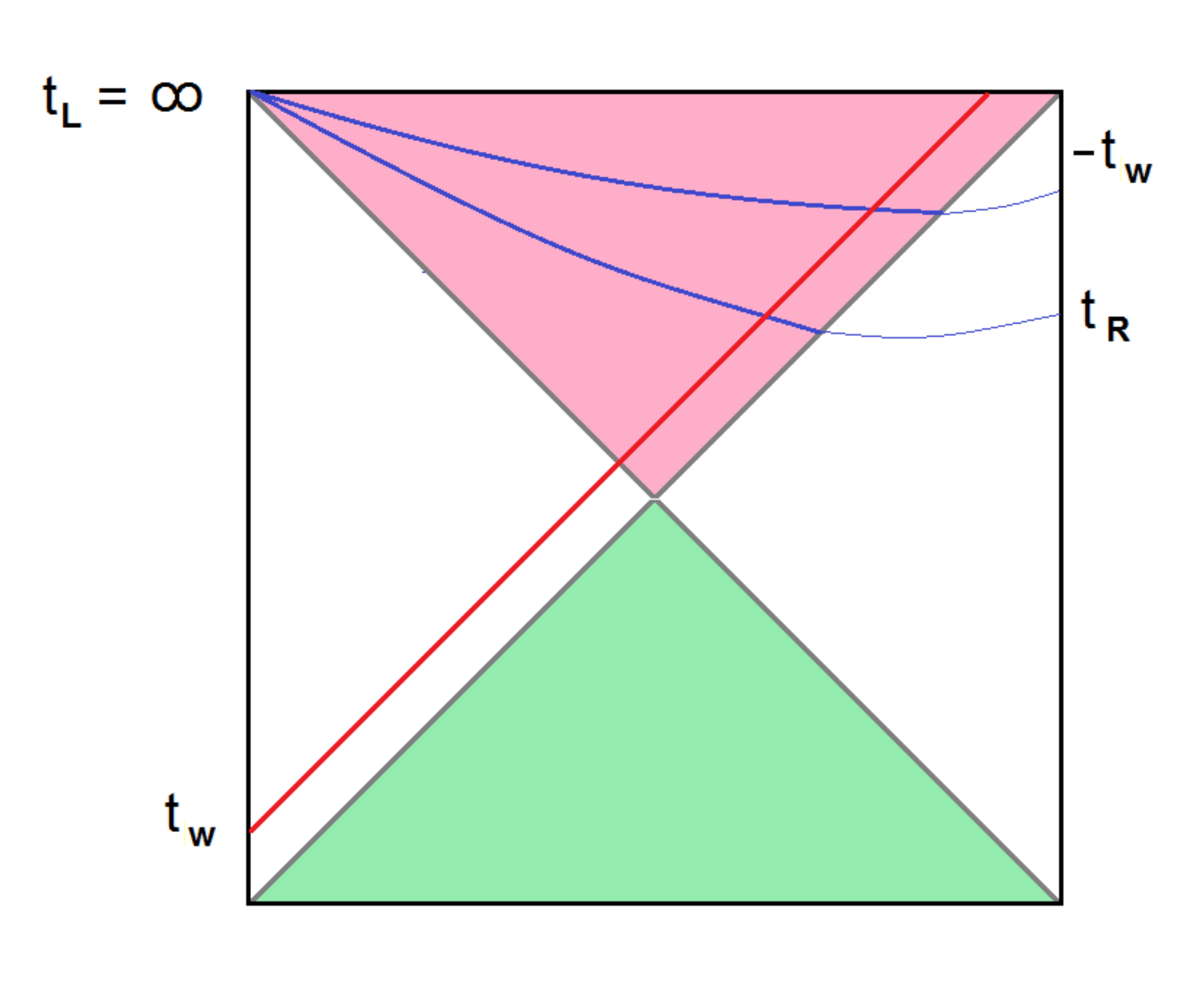}
\caption{A shock wave geomety with a shock injected on the left at time $t_w.$}
\label{fig4}
\end{center}
\end{figure}

Now let us consider a less trivial example in which the horizon is made opaque by a firewall in the form of 
 a Shenker-Stanford shock wave created by a computationally complex precursor operator \cite{Susskind:2014rva} that mimics a small perturbation on the left side     as in figure \ref{fig4}. The perturbation $W$ acts at time $t_w$ which we take to be negative and much larger than the scrambling time $t_{\ast}$,
\bea
t_w &<& 0 \cr 
|t_w| &>>& t_{\ast}.
\label{tw}
\eea 
The right horizon will  be opaque to an observer entering from the right, as long as the shock wave is within a Planck distance from the horizon. 

As before, if we want to diagnose the right horizon we hold the left anchoring point fixed at $t_L=\infty$ and vary the right anchoring time $t_R.$
In \cite{Stanford:2014jda} a simple formula was given for the volume of the ERB\footnote{Equation \ref{5.2} is actually the formula for the length of a geodesic anchored at the two boundaries. The actual formula for the volume of a maximal slice is more complicated, but as shown in \cite{Stanford:2014jda} it has all the same features as \ref{5.2} which provides a useful stand-in.} as  a function of $t_L$ and $t_R,$

\bn

\be
V(t_L, t_R) \sim  \log \left[ \cosh \frac{t_L+ t_R}{2}   + qe^{(2|t_w| +t_L -t_R)/2} \right]
\label{5.2}
\ee
with $q\sim e^{-t_{\ast}}.$ (The overall normalization in \ref{5.2} is irrelevant for our purposes.) Now take $t_L$ to be very large and define 

\be 
V(t_R) = V(t_L, t_R)-t_L/2.
\label{5.3}
\ee
we find,

 \be
V(t_R) = t_R/2  + \log{ \left[ 1 + e^{(|t_w| - t_R - t_{\ast})}\right]   }
\label{5.4}
 \ee

In \cite{Stanford:2014jda} the surprising fact  was noted that $V(t_R)$ is a \it decreasing \rm function of $t_R$ for 
$t_R < |t_w| - t_{\ast}.$ 
It was possible to connect the decrease with a reflection property  of maximally entangled states of the TFD type but here  we see a more far-reaching connection: opaque horizons---in this case  firewalls---are diagnosed by   decreasing complexity.
 
\begin{figure}[H]
\begin{center}
\includegraphics[scale=.6]{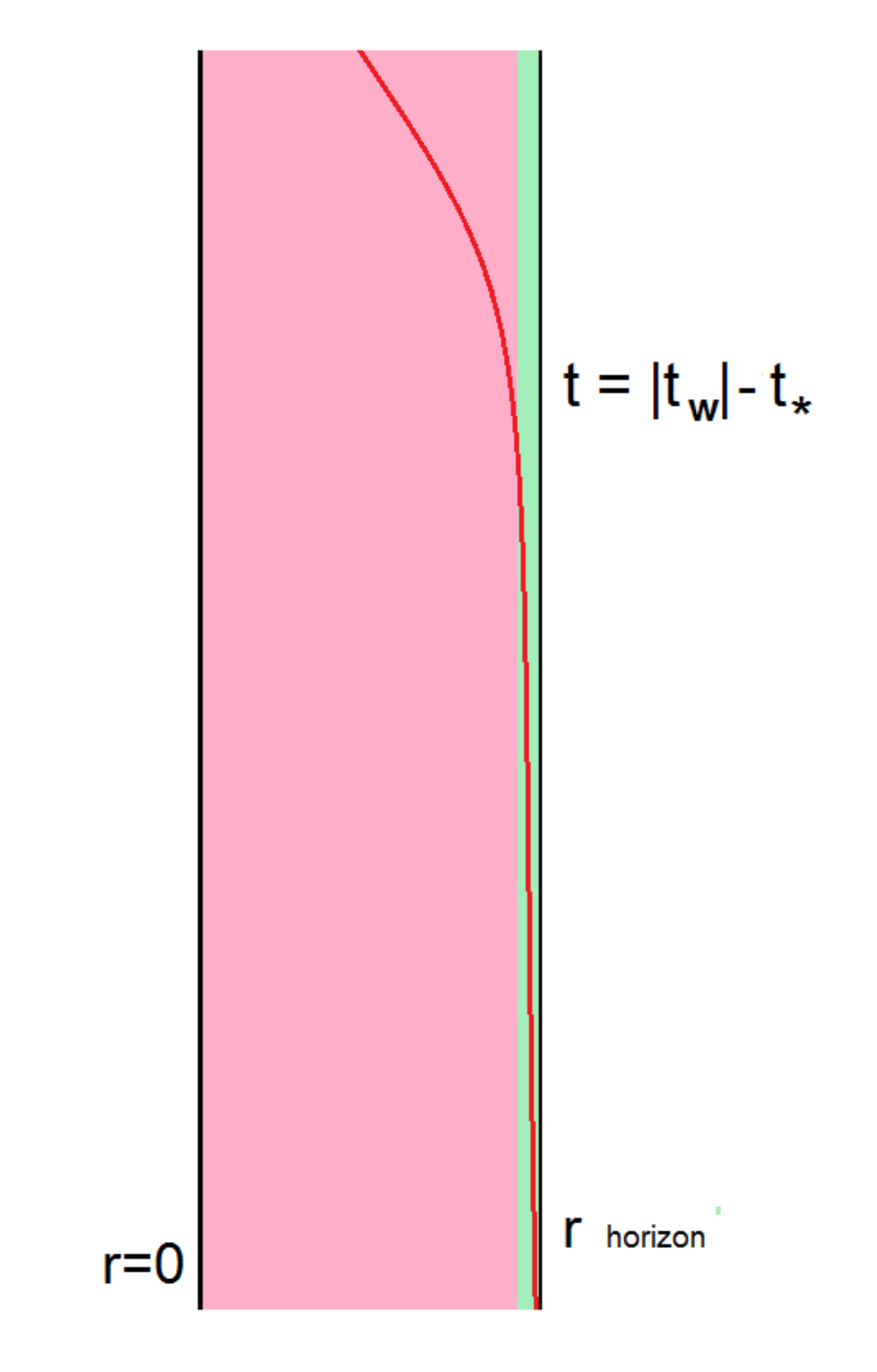}
\caption{Eddington-Finkelstein diagram for a shock wave initially within a Planck distance of the horizon. At the indicated time the shock separates from the horizon and falls into the singularity.}
\label{fig5}
\end{center}
\end{figure}

  In figure \ref{fig5} the evolution of the shock wave is shown in Eddington-Finkelstein coordinates. The light green region is within a Planck length of the horizon, and the red curve represents the shock wave. The shock wave remains in the Planckian region until time  $|t_w| - t_{\ast}$ when it splits off the horizon and falls into the singularity. The time at which it separates from the horizon is rather sharp, and it agrees exactly with the time at which the complexity stops decreasing. Thus we see that the complexity decreases precisely while the horizon is opaque.
  After that $V(t_R)$  and the complexity increase, and the horizon rather suddenly becomes \tr. 

The left horizon can be diagnosed  by holding $t_R = \infty$  and varying $t_L.$  From  \ref{5.2} one easily sees  that the volume of the ERB increases with increasing $t_L$. This implies that the left horizon is \tr.  

More can be done with multiple shocks using the solutions of  \cite{Shenker:2013pqa}. In the case of several shocks from the left side, the first shock to enter is the last shock to exit from the  green region in figure \ref{fig5}. It is easy to see that the ERB volume decreases until the last shock exits. The same pattern is seen for 
 shocks from both sides. 
All of this further supports the conjectured criterion for \try.  

It would be interesting to diagnose  the localized shocks of \cite{Roberts:2014isa}. The horizon in this case will become transparent in a non-uniform way with a wave of \try \ spreading out over a period of time\footnote{Dan Roberts has calculated the envelope of the spreading wave of transparency and finds that it spreads with the butterfly velocity \cite{Shenker:2014cwa}.}. I will leave further discussion of  this  for  a future paper.
\bn

\sc
\section{Very Long Times}

\begin{figure}[H]
\begin{center}
\includegraphics[scale=.3]{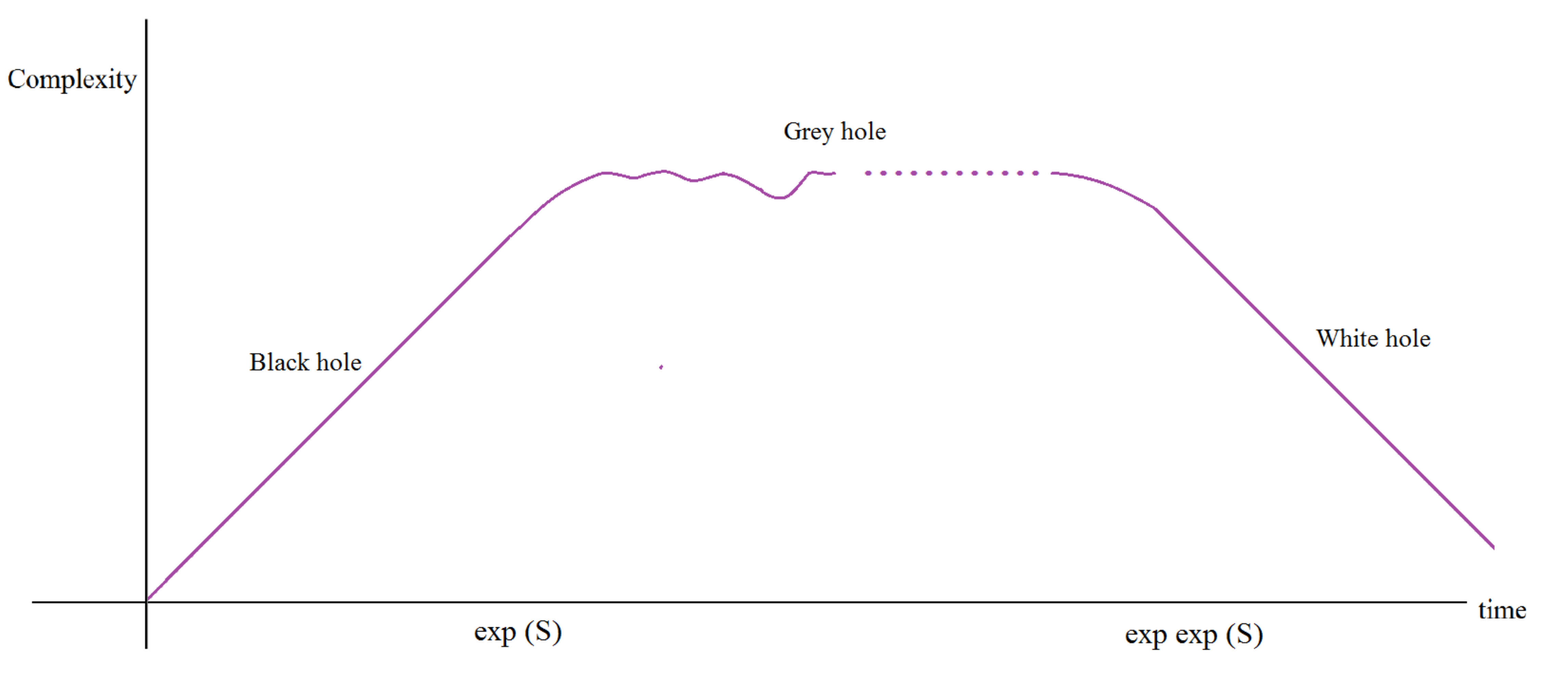}
\caption{Complexity history of a chaotic quantum system. The complexity begins low and increases for an exponentially long black hole era. It reaches complexity equilibrium which defines a ``grey hole" era. After a quantum recurrence time the complexity will quasi-periodically decrease during a white hole era.  }
\label{fig6}
\end{center}
\end{figure}

We would get into fewer paradoxes if we stopped calling things black holes when they  have a high component of white hole in their wave function. The ensemble of typical states is time-reversal symmetric with white holes being as likely as black holes. By contrast, natural black holes are objects that start out with low complexity, and evolve for a very long time before complexity reaches its maximum. They are black holes as long as the complexity  increases, after which they have a component of white hole behavior. In figure \ref{fig6} \  I've  sketched the complexity  history  beginning with a simple black hole $t=0.$ The actual behavior is much more irregular during the post-exponential period with fluctuations of many different amplitudes.
The expected behavior  is that  complexity increases linearly for an exponential length of time. This is the true black hole period.
At the other extreme, after a doubly exponential quantum recurrence time \ref{4.2}, the complexity will  become small again. Decreasing complexity defines the white hole period, during which the horizon is very vulnerable to instabilities. 

In between black and white hole behavior the system spends a time $\exp \exp (S)$ in complexity-equilibrium. During this ``grey hole" period there is no arrow of time to stabilize the geometry. It's not quite clear how horizons behave during this period, but following Marolf and Polchinski \cite{Marolf:2013dba}, I have assumed the worst; namely, grey hole horizons have a significant likelihood of being \op. This could be overly pessimistic.  

\bn

There is a puzzle about the evolution of the  exact thermofield-double state. The TFD is not time translation invariant, but it is boost invariant. For that reason the probability that an observer on the right side encounters an opaque obstruction is time-independent. If, as is widely believed, that probability is extremely small at $t=0,$ then it must be small for all time. On the other hand the complexity of the TFD is bounded by $\exp{S}.$ Therefore over  very long times there will be doubly exponential periods of complexity equilibrium, punctuated by singly exponential periods of white hole behavior. The white holes may not be a problem since they occur with a negligible probability of order $\exp{-\exp{S}}.$  But the doubly exponential period of grey hole behavior is more puzzling. If grey hole behavior means that the horizon is opaque, then at any time the probability to encounter a transparent horizon would be negligible.

There are a number of  ways to resolve the puzzle. One is that grey holes generally have a high probability of being transparent. Another is that the TFD is special in some way that gives extra protection to the transparent horizon. I will leave the resolution of this puzzle for the future.

\sc
\section{ The Fragility of Opacity}

Creating opacity is a computationally very expensive operation \cite{Susskind:2014rva}. Either it requires acting with a complex precursor or waiting a very long time until the black hole itself passes through exponentially many mutually orthogonal states. Once opacity has formed---by whatever means---it is very dangerous to an in-falling observer. But it is also very fragile. Opacity is like a very thin, hard,  brittle  shell at the horizon; if you crash into it it is deadly, but it is very easily broken. All it takes is a single thermal photon to destroy the shell and render the horizon transparent\footnote{This section was added after  helpful discussions with Tom Banks, Adam Brown, Brian Swingle, and Douglas Stanford. I am grateful to them for discussions about this material.}.

\bn

We have seen evidence that opacity and decreasing complexity are the same thing. One immediate consequence is that opacity is an extremely fragile condition. 
A state of decreasing complexity is very  unstable, rather like decreasing entropy.
Consider a fine-tuned state in which the complexity is on a decreasing trajectory. Such trajectories surely exist, they are the time-reversals of trajectories with increasing complexity. But disturbing a single low energy degree of freedom, say by dropping in a single thermal photon, will   quickly reverse the decrease. 

How long can decreasing complexity last after such a disturbance? To reverse the direction requires that the disturbance  effects a large fraction of the degrees of freedom. This is very closely related to the switchback effect of \cite{Susskind:2014jwa}\cite{Stanford:2014jda}. Therefore
 time for the disturbance to spread is the scrambling time. 
These are general facts about  any chaotic quantum system. When applied to black holes they imply that opacity is
unstable, and when disturbed will give way to \try \ in about a scrambling time. 

This complexity-theoretic analysis can be confirmed geometrically.
\begin{figure}[H]
\begin{center}
\includegraphics[scale=.3]{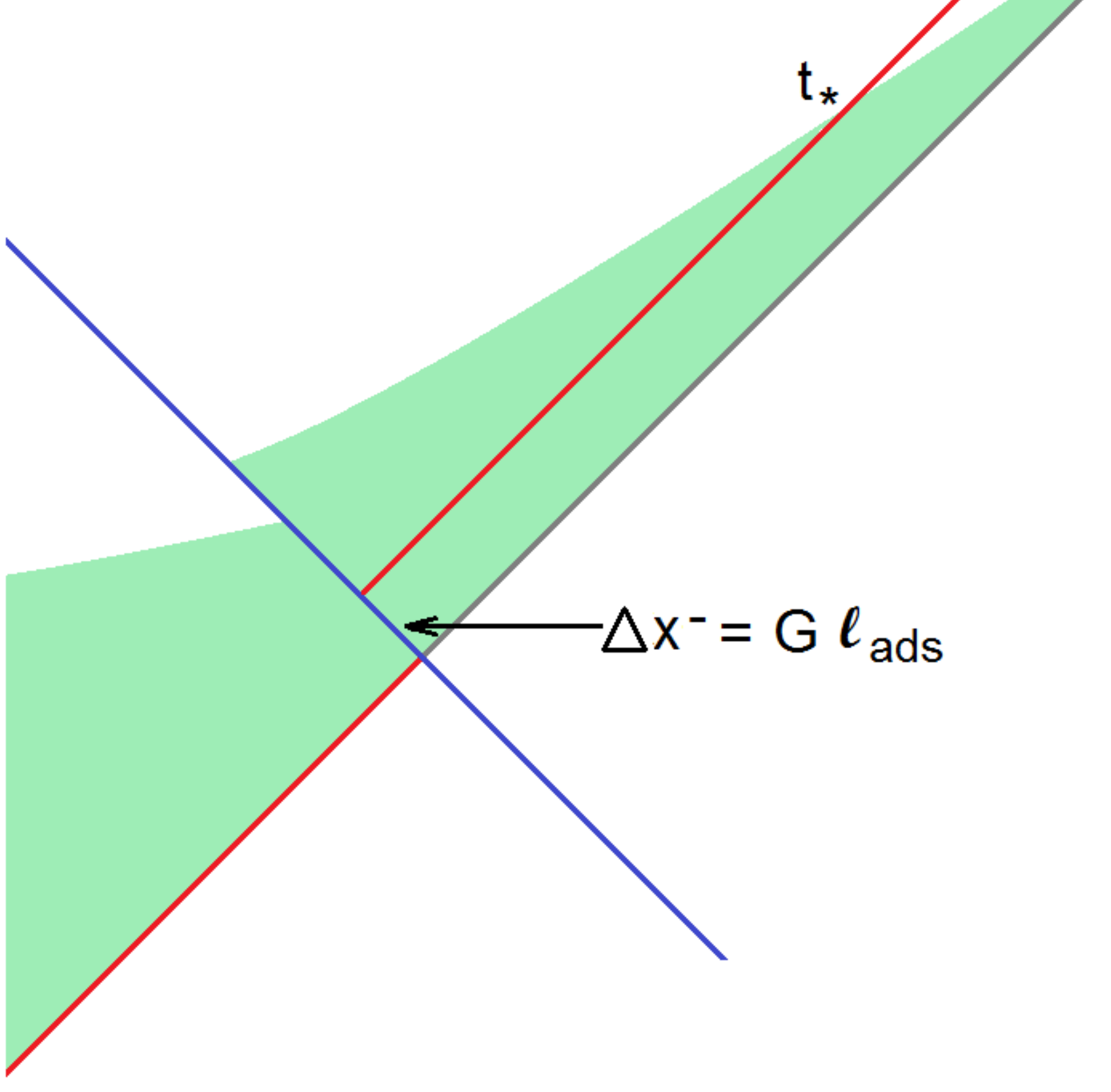}
\caption{A red shock wave propagating from the lower left encounters a blue photon. The collision displaces the shock wave into the interior by $\Delta x^-.$ The green portion of the diagram shows the region within a Planck length of the horizon. }
\label{fig7}
\end{center}
\end{figure}
\bn
Figure \ref{fig7}, drawn in Kruskal coordinates,  shows an opaque horizon   caused by a shock wave (red), along with a low energy photon (blue) falling on to it. The photon has energy $1/l_{ads}.$ The high energy red shock wave and blue photon intersect in a high energy collision. According to 't Hooft's description of such collisions \cite{Dray:1985yt}  the effect is to   shift the geometry  by amount $\Delta x^- = G l_{ads}. $ Equivalently the horizon moves outward in response to the additional energy.  The result is that the shock wave becomes displaced  relative to the horizon and proceeds to separate from it.  The shock wave  exits the (green) Planckian region in a scrambling time $t_{\ast},$ and  the horizon  becomes transparent. In another scrambling time the shock wave  will fall into the singularity.

Thus an observer who wants to enter a horizon and is not sure if it is transparent can clear it of any firewall by dropping in a thermal photon and waiting a couple of scrambling times.

What about grey holes; can they also be cleared as easily? Suppose we throw an extra thermal photon into a black hole that has come to complexity equilibrium. This will increase the entropy  by about one bit. More important, the exponential of the entropy will increase by a factor of two. This  means that the black hole will no longer be in complexity equilibrium. The complexity will be begin to increase and will continue to do so for another exponential time. According to the conjectured diagnostic the horizon will become transparent for a similar period.

To summarize, a firewall may indeed be a danger to an infalling observer. It forms an opaque, high energy shock wave that coats the horizon in a thin Planckian shell. But it is also a very fragile shell; the slightest perturbation will destroy it in a scrambling time. By contrast, the \try \ of a black hole on an complexity-increasing trajectory is extremely robust.

\sc
\section{Remark About Linearity}\label{S-linearity}

The typical state paradox may be resolved by the arguments in this paper, but even if this point of view is correct it 
 raises its own paradox of nonlinearity   
\cite{Marolf:2013dba}\cite{Bousso}.   
 One might expect that the property of being a black hole, as opposed to a white or grey hole, should define a linear subspace of the Hilbert space. But it is clear that  the states on the rising part of the complexity curve are a complete set. It is even more clear that the states on the complexity equilibrium part of the curve are vastly over-complete. It appears that  the span of the black hole states is the full Hilbert space: the span of the white hole states is the full Hilbert space: and the span of the grey hole states is the full Hilbert space.  Evidently the property of being a black hole cannot be a conventional observable. 
The point is reinforced by the observation that complexity is not a linear property. The superposition of two low complexity states will generally be of high complexity. 

I will leave this paradox for a separate paper.

\section*{Acknowledgements}
I am grateful to Tom Banks, Adam Brown, Dan Roberts, Steve Shenker and  Douglas Stanford for helpful discussions about shock waves and the fragility of opacity. 

This work was supported in
part by National Science Foundation grant 0756174 and by a grant from the John Templeton Foundation.
The opinions expressed in this publication are those of the author and do not necessarily
reflect the views of the John Templeton Foundation.

\end{document}